# All-optical and ultrafast control of high-order exciton-polariton orbital modes

Yuyang Zhang[1,2#], Xin Zeng[1,3#], Wenna Du[1,2#], Zhiyong Zhang[1,2], Yuexing Xia[1,2], Jiepeng Song[4], Jianhui Fu[1,2], Shuai Zhang[5], Yangguang Zhong[1,2], Yubo Tian[1,2], Yiyang Gong[1,2], Shuai Yue[1,2], Yuanyuan Zheng[1,2], Xiaotian Bao[1,2], Yutong Zhang[1,2], Qing Zhang[4*], Xinfeng Liu[1,2*]

[1]CAS Key Laboratory of Standardization and Measurement for Nanotechnology, National Center for Nanoscience and Technology, Beijing 100190, China

[2]University of Chinese Academy of Sciences, Beijing, 100049, China

[3]Tianjin Key Lab for Rare Earth Materials and Applications, Smart Sensing Interdisciplinary Science Center, Nankai University, 300350, Tianjin, China

[4]School of Materials Science and Engineering, Peking University, Beijing 100871, China

[5]School of Physics and Electronic Information, Weifang University, Weifang, 261061, China

*Email address: liuxf@nanoctr.cn, q_zhang@pku.edu.cn

**Exciton-polaritons flows within closed quantum circuits can spontaneously form phase-locked modes that carry orbital angular momentum (OAM). With its infinite set of angular momentum quantum numbers ($\hbar$), high-order OAM represents a transformative solution to the bandwidth bottleneck in multiplexed optical communication. However, its practical application is hindered by the limited choice of materials which in general requires cryogenic temperatures and the reliance on mechanical switching. In this work, we achieve stable and high-order (up to order of 33) OAM modes by constructing a closed quantum circuit using the halide perovskite microcavities at room temperature. By controlling the spatial and temporal symmetry of the closed quantum circuits using another laser pulse, we achieve significant tuning OAM of EP flows from $8\hbar$ to $12\hbar$. Our work demonstrate all-optical and ultrafast control of high-order OAM using exciton-polariton condensates in perovskite**

**microcavities that would have important applications in high-throughput optical communications.**

Orbital angular momentum (OAM), an additional spatial degree of freedom for beams, has recently garnered significant interest for its potential to overcome the bandwidth limitations in optical information transmission[1–4]. High-order OAM modes with a large number of $\hbar$ (Plank's constant) which carriers more information, are thus well-suited for applications like multiplexing in optical communication[5–9]. Such high-order OAM can usually be generated in a closed coherent quantum circuits due to the quantum wavefunction coupling and spontaneous phase-locking under periodic boundary conditions. These coherent systems can be created by quasi-lattices[10], microring resonators[11–13], superfluid systems[14,15] and etc.

Among those methods, exciton-polaritons (EPs) are ideal candidates for constructing such coherent systems[16]. This is because EPs are formed through the strong coupling between photons and excitons, thereby possessing properties from both photon and matter[17]. The photon component endows EPs with an exceptionally small effective mass, allowing them to exhibit Bose-Einstein condensates (BEC)-like behaviors[18–21], and form condensation without the need for the cryogenic conditions required in cold atomic systems[22–28]. The resulting superfluid effects create an effective platform for simulating quantum coherent systems[29–31]. Simultaneously, the exciton component introduces strong nonlinear interactions, enabling precise manipulation of EPs flows using external potentials[32–37]. By geometrically shaping the photon mode or light field into a circular distribution, the EPs flows can spontaneously rotate[38]. This counterrotation leads to vortex-antivortex superposition, forming phase-locked condensates, which manifests as unique petal-shaped emissions. Such petal-shaped emissions signify the generation of OAM modes and exhibit remarkable stability under continuous pumping[39–41].

Initially observed at liquid-helium tempertures[42], these emissions can be controlled by varying factors such as the radius of the potential barrier[43], the dimensions of etched microstructures[44–47], the intensity distribution of the light field[48] and etc. Compared with these methods, all-optical control introduces a versatile approach by enabling in-situ, rapid, and widely adjustable OAM operations using both spatial and temporal degrees of freedom[49,50]. Furthermore, the uniform photon mode in this method ensures pure excitation modes, eliminating the influence of additional modes on OAM formation[51]. On the other hand, low-order OAM modulations typically exhibits a weak response, which often lack the ability to deliver the continuous and efficient control needed for practical applications. Therefore, a clear understanding of high-order OAM behavior as well as an efficient control method is key to developing OAM-based technologies. Moreover, emerging perovskite materials possess large exciton binding energy and giant oscillator strength[52,53]. When used as gain media in microcavity systems, they facilitate the formation of stable quasi-BEC condensation and long-range coherent EPs flows at room temperature[54–59]. This development addresses the constraints of conventional systems, which typically require low-temperature operation, thereby greatly improving their practicality and expanding their application potential.

In this work, we design an annular optical potential barrier within a planar microcavity, consisting of single crystalline $CsPbBr_3$ nanosheets, silver (Ag) and a distributed Bragg reflector (DBR). By using a ring-shaped above-bandgap pump, we create a cylindrically symmetric potential barrier which enables us to generate a high-order OAM mode with a petal-shaped coherent emission at room temperature. Phase analysis using the Mach-Zehnder interference spectroscopy reveals a π-phase interval between the petal structures, indicating the presence of vortex-antivortex pairs within the condensates. Moreover, by controlling the spatial and temporal symmetry of the potential barrier using another femtosecond laser pulse, we achieve significant in-situ orbital mode tuning of the OAM modes. Such efficient modulation is enabled by the strong nonlinear and dynamics interactions between EPs, coupled with the strong response of high-order OAM. Our work demonstrates forming and ultrafast control of high-order OAM modes by leveraging the EPs condensates in perovskite microcavities, which would be important not only for advancing the study of coherent quantum circuits but also lay the foundation for developing novel information encoding technologies, such as polarization devices, optical cryptography, and optical sensing, etc.

**Results**

**Fig. 1a** illustrates the schematic of the perovskite microcavities which consist of Ag layer, $CsPbBr_3$ nanosheet layer and the DBR substrate, forming the Fabry-Pérot (F-P) microcavity. The $CsPbBr_3$ nanosheets are grown on the DBR substrates using an anti-solvent and template-assisting method. Details of the sample preparation are shown in **Supplementary Fig. 1**. To generate high-order OAM modes, we artificially create the periodic boundary conditions for EPs condensates by employing a ring-shaped pulse which is generated by shaping the 400 nm laser pulse using a 0.5° axicon. The ring-shaped pulse serves as a pump source to excite in-situ excitons, establishing strong coupling with cavity modes to form EPs. The repulsive interactions between locally accumulated excitons and EPs create a potential barrier at the position. Due to the accumulation of excitons and EPs excited by the ring-shaped pulse, fluorescence emission overlapping with the excitation ring pattern can be observed under pumped below the threshold. The cylindrical symmetry of the excitation and the annular potential barrier cause the EPs flows spontaneously rotate along the inner edge. As the excitation power increases, nonlinear effects become more pronounced, leading to the condensation behavior. When the excitation power reaches the threshold, the emission intensity increases sharply, and the vortex-antivortex superimposed EPs flows form a special petal-shaped pattern in real space **(Fig. 1b)**. The petal-shaped condensates preserves some of the system's cylindrical symmetry. The angle-resolved photoluminescence (ARPL) spectra of EPs also change with the power increase, switching from fluorescence distribution below the threshold (left side) to the condensation distribution above the threshold (right side), as shown in **Fig. 1c**. The lower polariton branch (LPB) exhibits an anticrossing dispersion, confirming the realization of strong exciton–photon coupling (the contrasts with the pure Fabry-Pérot (F-P) modes are shown in **Supplementary Fig. 3**). The generation of petal-shaped condensation can be verified at different positions and circular radii of the potential barriers in the experiment, demonstrating the the repeatability of the system. By adjusting the size of the circular pulses and the detuning of the samples, the number of petals can be controlled. The petal number and the size of the potential barrier are positively correlated ($n \propto R^2$), where n is the petal number and $R$ is the radius of the ring barrier. We achieve condensates with varying petal numbers **(Supplementary Fig. 5)**, with the maximum petal number of 66 realized in the experiment. As shown in **Fig. 1b**, the number of lobes remains even, following the rule of n = $2l$, where $l$ is the Laguerre-Gauss (LG) azimuthal mode. The generation of petal-shaped patterns arises from the superposition of EPs flows, specifically the combination of two Laguerre-Gauss modes ($\pm l$). The counterrotating superposition can be demonstrated through far-field k-space imaging through a bidirectionally adjustable rectangular slit, which allowed selection of different numbers of lobes. Each independent lobe selected in the far field results from the superposition of two opposite wave vectors, confirming the presence of counterrotating EPs flows. As the number of lobes in the selected region increases, the occupation of the wave vector in the ky direction has expanded. When more than half of the lobes are selected, reciprocal petal-shaped patterns emerge in far-field imaging, providing strong evidence for the existence of the ± $l$ LG mode superposition **(Supplementary Fig. 6)**.

Furthermore, we employ M-Z interferometer spectroscopy to analyze the mechanism of petal-shaped condensation through phase analysis **(Supplementary Fig. 7)**. Interference fringe patterns of the condensates indicate a good coherence of the petal-shaped emission **(Fig. 1d)**. By adjusting the distance between the M-Z arms, we can monitor the evolution of interference patterns. Clear fringes are shown within 5 ps delay, verifying the long-range coherence of the condensates. Notably, interference fringes between adjacent lobes exhibit 'Y'-shaped dislocations, representing phase jumps in boundary regions. After extracting the interference pattern numbers using off-axis holographic technology[60], we obtained the phase distribution of the petal-shaped condensation **(Fig. 1e)**. The phase distribution exhibits clear partitioning behavior, and the height corresponds to the number of lobes in Fig. 1d. The phase distribution reveals the uniform phase distribution inside each lobe and the π - level phase transition between adjacent petals. High-order emission causes slight mutual influence on the lobe edge, but still maintains basic emission properties. Under the excitation of cylindrical symmetry, the petal-shaped condensation distribution exhibits a cross distribution along the angular orientation. Strictly speaking, in a perfectly uniform cylindrical symmetric potential barrier distribution, the LG modes with zero radial exponent will linearly superimpose, and the energy levels should be 2-fold degenerate, appearing as two opposing points on the equatorial x-axis on the Poincaré sphere. At this point, a bright green ring, shaped like the pulse excitation, should appear at the threshold instead of the petal-shaped condensation. However, in practice, factors such as the pump distribution, sample conditions, and the presence of defects introduce additional perturbations such that the counterrotating LG modes couple in energy separation, resulting in excitation collapse into petal-shaped distributions and the formation of spontaneously phase-locked condensates. As the excitation power continues to increase, due to the expansion of the Gaussian-distributed boundary, the effective area of the circuit decreases, and the mutual influence between lobes increases. The petal-shaped emission gradually expands and merges until it evolves into a singular vortex state. This phenomenon is caused by the dynamic competition between excitons and EPs. As a result, the petal-shaped condensation only exists near specific excitation thresholds. In the phase distribution, some additional vortex behaviors at different radial positions also appear inside the annular barrier **(Fig. 1e)**, which illustrates the LG emission with $p \neq 0$ in the case of large spatial expansion, further confirming the agreement between experiments and theory. This petal-shaped emission serves as a macroscopic and intuitive representation of OAM emission. Rapid and linear control of this state is extremely important for the study of quantum circuits and photon cryptography.

To achieve ultrafast control of orbital modes, we introduce additional control pump pulse with relative delay time of 0 ps. The position of the control pulse relative to the center of the ring pulse changes the cylindrical symmetry of the EPs **(Fig. 2a)**. The closed quantum circuits generated by the annular optical barrier can achieve free control of orbital modes by introducing all-optical modulation with spatial and temporal degrees of freedom. The spatial degrees of freedom are reflected in the response of phase distribution change with different geometric symmetries. **Figure 2b** represents the potential barrier distribution and the change of polarization after adding control pulses at the ring. At this point, controlling the additional potential barrier introduced by pulses can enhance the EPs interaction in the region, thereby affecting the distribution of orbital modes. The control pulse can strongly affect the distribution of petals nearby, while those farther away occur slightly changing. This is attributed to the dependence of the coupled orbital mode distribution and the barrier radius, ensuring the robustness of the petal distribution in the indirect action region. The local nonlinear strength change obtains the highest response near the direct acting region, and the orbital mode switches accordingly. Here, we precisely quantify the control effect of potential barrier symmetry on emission by focusing on the interference phase changes of petals in a fixed area. For clarity, we excited the initial petal distribution through ring pulse and selected the area with the most uniform petal phase and highest resolution for focused observation and controlling. As a result, we chose a region with 11 lobes at the bottom of the ring as the focus area, the phase change of the area can be distinguished after phase analysis (**Fig. 2d**). By further introducing a cylindrical symmetric intensity distribution control pulse placed in the middle, slight enhancement of petal emission and expansion of phase distribution can be observed in the real image, corresponding to an increase in emission intensity. When the control pulse is located outside the center, the cylindrical symmetry of the system is broken, and the shape of the quantum circulation is distorted. The original uniform petal emission degrades, and the mode changes with the symmetry breaking. By adjusting the position of the control pulse, the strongest modulation effect is obtained at the position in contact with the annular barrier (**Fig. 2c**). The phase distribution expands significantly, with the original uniform emission of 11 modes switching to 9.3 modes. The decimal mode count results from incomplete petals within the fixed area, reflecting the expansion of the coupled orbital mode. By modulating geometric symmetry on high-order OAM systems, rapid and significant control over OAM behaviors can be achieved, which is difficult to observe and quantify in traditional low-order OAM systems. Due to the correlation between the number of lobes and the size of the potential barrier ($n \propto R^2$), more efficient modulation systems can be predicted in a higher order system.

To further enhance the capability of all-optical modulation, we investigate time-resolved photoluminescence dynamics of the EPs flows. Building on the control of symmetry described above, time-resolved non-resonant excitation is performed at the most effective position, where the control pulse overlaps with the ring pulse **(Fig. 3a)**. The delay time between the two beams is achieved by adjusting the position of the delay line **(Supplementary Fig. 2)**. The different time positions of the control beam represent the different processes in which the control pulse acts as an additional potential barrier to participate in the establishment of circulation dynamics, thereby modulating the gain-loss mechanism of petal-shaped emission and allowing for the free manipulation of the orbital modes. **Figure. 3b** shows the corresponding dynamic process. Due to the continuous adjustability of time regulation, multiple orbital modes can be freely selected within the regulation range, demonstrating a high degree of freedom in tuning of high-order OAM systems. In the experiment, we demonstrated that significant changes in petal-shaped emission were observed in real-space image and phase analysis by adjusting the delay time between of two pulses. In the phase map of the focusing region, the emission modes of 8-11 were freely controlled. **Figure 3c** shows the variation of petal shaped partition in phase analysis with the changes of delay time. $\Delta t_{AB}$ represents the delay time of the two pulses. As the delay time of the two pulses increases, the control pulse participates in different dynamic processes of petal-shaped condensation: P1 (thermal excitons state), P2 (formation of EPs), P3 (EPs relaxation), P4 (EPs condensation at the bottom of the LP band), as shown in **Fig. 3b**. The regulation of different dynamics brings corresponding differences in the number of orbital modes. **Figure 3d** highlights the detailed evolution of localized modes throughout the entire process, providing deeper insights into the mechanisms involved. When the controlling pulse reaches before coherent oscillation is formed (t < 0 ps), it provides an extra exciton reservoir, which also serves as an additional potential barrier, forming a low number of modes corresponding to the P1 process. The reverse trend of modulation enhancement near zero-point time (0 ps < t < 7 ps) corresponds to the short density saturation of absorbing states caused by the initial ring pulse excitation. The absorption of the control pulse recovere at 7 ps, at which point the control effect comes to strongest. Other factors can also contribute to the occurrence of this phenomenon, such as the delayed establishment process of petal condensation, or the exciton diffusion of control pulse, corresponding to the lowest mode number in the phase distribution **(Fig. 4c)**. It is worth noting that the control effect of the additional potential barrier will generate a linear response region (7 ps < t < 33 ps) in the P3 process. This process can be clearly distinguished by changing and expanding the number of petals in the focusing region. Switching speed at the picosecond level also proves the potential application of temporal freedom in all-optical control. As the time scale further increases (t > 33 ps), petal-shaped condensation tends to be stable and the coherence between the polaritons generated by the ring pulse and the additional potential barrier disappeares. The resulting petal-shaped condensation becomes essentially invariant, and increasing the power of control pulse only increases local fluorescence emission without changing the phase distribution and coherence **(Supplementary Fig. 8)**. Furthermore, by adjusting the radius of the annular potential barrier, the intensity of the pulse, or the detuning of the cavity, the range of the modes can be freely configured, which provides important reference for OAM applications at room temperature.

As a potential option for the development of optical communication, the robustness of the system serves as a crucial indicator for practical applications. In this study, we verified the stability of petal-shaped emission by sequentially opening the closed annular barriers. Initially, the energy level of condensation generated by the fully closed annular barrier is dominated by $E = 2.304$ eV, exhibiting strong and stable emission with a high fringe visibility (FV = 0.74), as shown in **Fig. 4a**. **Figures 4b-4e** illustrate the changes happened on the real image surface as the barrier gap gradually opens. As shown, the dominant mode can remain stable within the gap range of 0° to 45°. The slight fluctuation may be due to the dark soliton behavior with the dominated energy level of the petal-shaped emission remaining unchanged. At this point, the condensation superfluid keeps robustness through a mechanism similar to Josephson junctions. As the gap continues to widen, the circulation fully opens, and the coherent mode will expand outward, corresponding to the orbital modes with a larger radius. The ARPL spectra in **Fig. 4f-4j** also show energy level switching, where the condensation energy band is peaked at 2.306 eV. When the gap expands to 120°, the annular barrier is insufficient to satisfies the requirements of a closed quantum circuit, and the emission mode inside the barrier transitions to a chaotic turbulence-like distribution. As the gap increases to 150°, the overall excitation fails to meet the requirements of EPs condensation. Instead, the PL imagings become arc-shape and the ARPL spectra is characterized by the fluorescence distribution with with a larger linewidth. The linear gap modulation verifies the emission stability of the system and confirms the feasibility for practical optical communication applications at room temperature.

**Discussion**

We demonstrate effective control of high-order OAM based on the nonlinear interaction of EPs. The additional optical potential barriers influences the superposition and phase-locking of the counterrotating EPs flows, resulting in modulation of coupled OAM modes. Based on the correlation between order and response effect, we achieve up to 33 order OAM modes in perovskite microcavities, thereby achieving efficient OAM control. The angle-resolved photoluminescence, far-field image and phase analysis of the condensates verifies the generation of high-order OAM modes. Leveraging the high order and the strong nonlinearity, the all-optical control shows a high control range (from $8\hbar$ to $12\hbar$) and ultrafast response. Our strategy is effective in both symmetric and temporal control, thanks to the freedom endowed by all-optical generation. The coupling of positive and negative OAM modes demonstrate sensitivity to additional potential barriers and robustness to gap variations. This asymmetric response aligns with practical application requirements for high responsiveness and repeatability. Combined with the room temperature and in-situ working conditions enabled by perovskite microcavities, our work lays the foundation for developing novel information encoding technologies. The small volume and ultrafast response of each cell may also pave the way for the advancement of integrated optical communication.

**Methods**

**Preparation of PDMS:** PDMS master mold was produced using electronic etching, UV exposure and high-density plasma silicon etching. Firstly, a positive photoresist is coated on a four-inch silicon wafer. Then, a designed photomask was placed onto the wafer, with the pattern of holes on the photomask being created through CAD software design and halogen silver development and exposure processing. Afterwards, UV exposure (Ma-6, Germany, SUSS) and silicon etching (ICPRIE-180, UK, Oxford Instruments) were performed to transfer the designed pattern onto the target PDMS silicon master, resulting in a silicon array with circular holes of 300 nm depth, 1 mm period, and diameters ranging from 0.25 to 0.8 mm. The master mold was sequentially cleaned with acetone, isopropanol, and ethanol. It was then immersed in a solution of 0.5 mM FDTS in heptane for 3 minutes to achieve hydrophobic treatment, which facilitates subsequent demolding. Mix PDMS (SYLGARD 184) base material and curing agent in a ratio of 10:1, stir evenly, pour onto a silicon master mold, vacuum treat for 1 hour to remove bubbles in PDMS, and then cure at 100 °C for 1 hour. After that, the target template is carefully removed from the master mold. Each surface hydrophobic treatment can support the production of more than 10 PDMS templates without the need for cleaning and without damaging the silicon master plate.

**Sample preparation:** DBR mirrors were fabricated on a silicon substrate using electron beam evaporation method (19.5 pairs of $TiO_2$ and $SiO_2$ with thicknesses of 54 and 88 nm). A precursor solution with a concentration of 0.4 M was prepared by dissolving stoichiometric ratios of CsBr and $PbBr_2$ powders (99.999 %, Sigma Aldrich) in DMSO. The solution was filtered and set aside for use. Ultrasonic treatment was performed on a 1 $cm^2$ DBR substrate in acetone and ethanol for 5 minutes, then dried and treated with oxygen plasma to make the surface hydrophilic. The substrate was then placed on a 3 cm high column, and 10 ul of precursor solution was dropped onto the DBR surface. The segmented PDMS template was then placed on top and subjected to weak pressure for 30 minutes before removal. Transfer the column to a 100 ml beaker, add 4 ml of acetonitrile solution to the beaker, seal it and keep it on a hot table at 38 °C for 24 h. The acetonitrile vapor environment can effectively separate perovskite and water oxygen. Finally, remove the column, peel off the PDMS template, and complete the sample growth.

**Microcavity fabrication:** PMMA particles (Polymethyl methacrylate, P821346-100g, MACKLIN Inc.) were dissolved in a toluene solution to prepare a 20 mg/ml precursor. After the sample growth was completed, 60 mg of the precursor solution was spin-coated at 5000 rpm for 60 seconds. The sample was then annealed at 90 °C for 5 minutes to form the protective layer. Afterwards, a 40 nm Ag film was evaporated on the top using a high vacuum evaporation coating system (OHMIKER-50B, Chongwen Technology), and a PMMA solution was spin coated again in the same way as a protective layer to slow down Ag oxidation, resulting in a complete Ag DBR cavity.

**Sample characterizations:** Atomic force microscope images were obtained using a BRUKER Dimension Icon. Steady-state fluorescence was obtained by focusing a 405 nm CW laser on a 100x objective lens (Olympus, ×100, NA = 0.9) and collecting it using a liquid nitrogen-cooled spectrometer (iHR-550, Horiba). The fluorescence spectra were obtained by grating spectroscopy at 600nm. Steady state fluorescence was obtained by focusing a 633 nm CW laser with a 100 × objective lens (Olympus, × 100, NA = 0.9), and the reflected Raman signal was collected by Horiba

JY T64000 and analyzed with an 1800 nm grating to obtain the Raman spectra. Angle-resolved spectra were obtained using a custom-built 4 f Fourier imaging system. Angular information was collected through a 50 × objective lens (Olympus, × 50, NA = 0.8), and an angular resolution of ± 53 ° was obtained through slit splitting in the Fourier phase plane. The signal was collected through a liquid nitrogen-cooled spectrometer (iHR-550, Horiba), and analyzed with 150 nm and 600 nm gratings to obtain momentum-resolved absorption and photoluminescence (PL) spectra. The white light source for the absorption spectra comes from a tungsten halogen lamp source (SLS201L, Thorlabs), and the 400nm laser source is obtained by doubling the frequency of an 800nm pulse (80 fs, 1 kHz, Astrella, Coherent) through a BBO crystal. The 400nm pump beam was split into two beams using a beam splitter, then a dual beam time-space modulation system was constructed with a delay line. The detailed optical setup is illustrated in **Supplementary Fig. 2**.

**Mach-Zehnder interferometer:** Coherence testing was obtained using a custom-built Mach-Zehnder (M-Z) interferometer. In the Fourier system, the real image plane was captured, and then a beamsplitter divided the image into two beams, with one beam passing through a lens for magnification and interference with the other beam. An adjustable delay line allowed for testing of coherence capabilities. The interference patterns were captured in real time using a grayscale camera (PL-D775MU-T, Pixelink).

**Phase Extraction:** Phase distribution was extracted using digital off-axis holography. By performing Fourier transform (FFT) on the interfered pattern to obtain continuous terms in the central region and interference terms in the off-axis region, filtering out the continuous terms, and taking one of the interference terms for inverse Fourier transform (iFFT), the distribution of phase and intensity of the original emission can be analyzed. The regions of interest in the phase map are extracted and analyzed uniformly using MATLAB.

**Associated content**
[*]Supplementary Information
The Supplementary Information is available.
Supplementary Fig. 1-8 as described in the text.

**Author information**
Corresponding Authors

*E-mail: liuxf@nanoctr.cn, q_zhang@pku.edu.cn

**Author contributions:** X.L. led the project. X. L. and Q. Z conceived the idea. X. L. and Y.Z. designed experiments. Y.Z., X.Z., S.Z. and Z.Z. set up the angle-resolved photoluminescence optical paths. Y.Z., Y.X., W.D. and X.Z. conducted the angle-resolved photoluminescence experiments. Y.Z., X.Z., S.Z. and Y.Z. grew the samples. S.Z. and W.D. designed the optical cavities. W.D., J.S., Y.T. and Y.G. led the experimental analysis. S.Y., J.F., Y.Z., X.B. and Y.Z. prepared the manuscript. All authors discussed the results and revised the manuscript. Y.Z., X.Z. and W.D. contributed equally to this work.

**Acknowledgments**

**Funding:** We thank the funding support from the National Key Research and Development Program (2023YFA1507002), the National Science Foundation for Distinguished Young Scholars of China (22325301), the Strategic Priority Research Program of Chinese Academy of Sciences，Grant No. XDB36000000, the National Natural Science Foundation of China (22073022, 12074086, and 22173025), and the CAS Instrument Development Project (Y950291). We also thank the funding support from the Natural Science Foundation of Beijing Municipality (JQ21004), Natural Science Foundation of Shandong Province (Grant No. ZR2024QA131)，The Young Scientists Fund of the National Natural Science Foundation of China (Grant No. 12204123)and the National Natural Science Foundation of China (52072006 and U23A2076). The authors thank the processing instrument support from NCNST：Mrs. L Yan, Mr. Z Song, Mrs. L Xu.

**Competing interests:** Authors declare no competing interests.

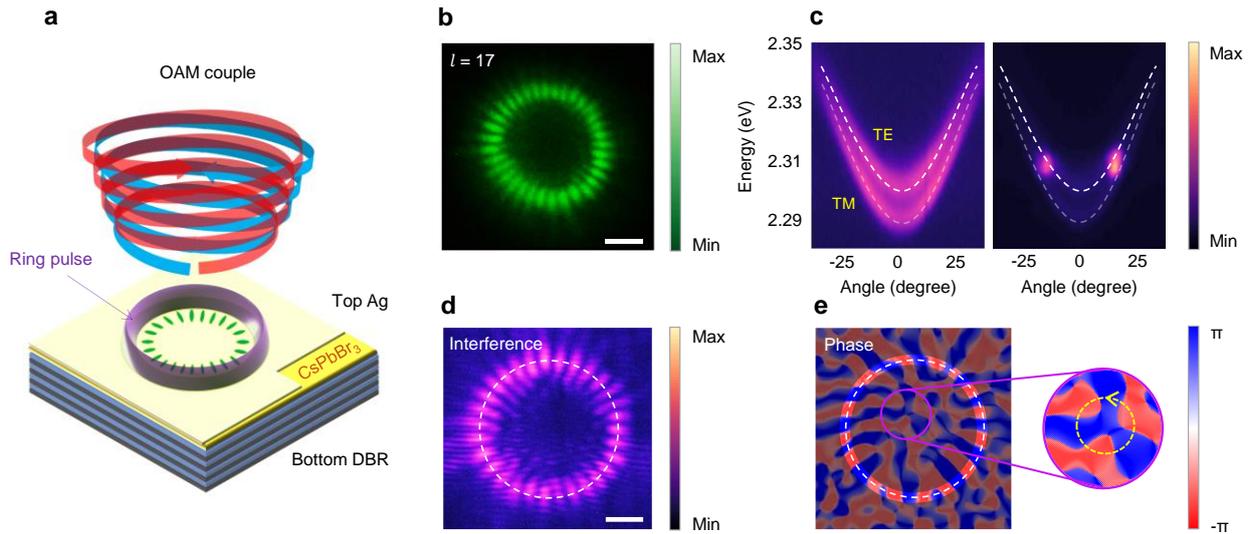

**Figure 1| High-order EPs orbital modes in perovskite microcavities.** (**a**) Schematic of coupled OAM modes generated by EPs flows in perovskite microcavity. The annular optical potential well induces the generation of coupled orbital modes and forms petal-shaped condensation at room temperature. (**b**) High-order petal-shaped condensates in real-space images. Corresponding to orbital modes with $l$ = 17 and n = 34. Scale bar is 5 μm. (**c**) Experimental momentum-space spectra of condensation. The emission below the threshold exhibits a fluorescence distribution, with the energy spread across the entire LPB (left part). In contrast, the emission at the threshold demonstrates a condensate-like distribution, where the energy is concentrated in specific modes corresponding to different LG modes (right part). The white dashed lines indicate the splitting of the TE and TM modes. (**d**) The interference of uniform petal-shaped condensates obtained by the Mach-Zehnder interferometer. Scale bar is 4 μm. (**e**) Phase distribution of petal-shaped condensates. The interference fringes in (**d**) were extracted through digital off-axis holography, and the highlighted rings represent the positions of the petals, proving the π phase shift between lobes and the radial secondary orbital modes behavior.

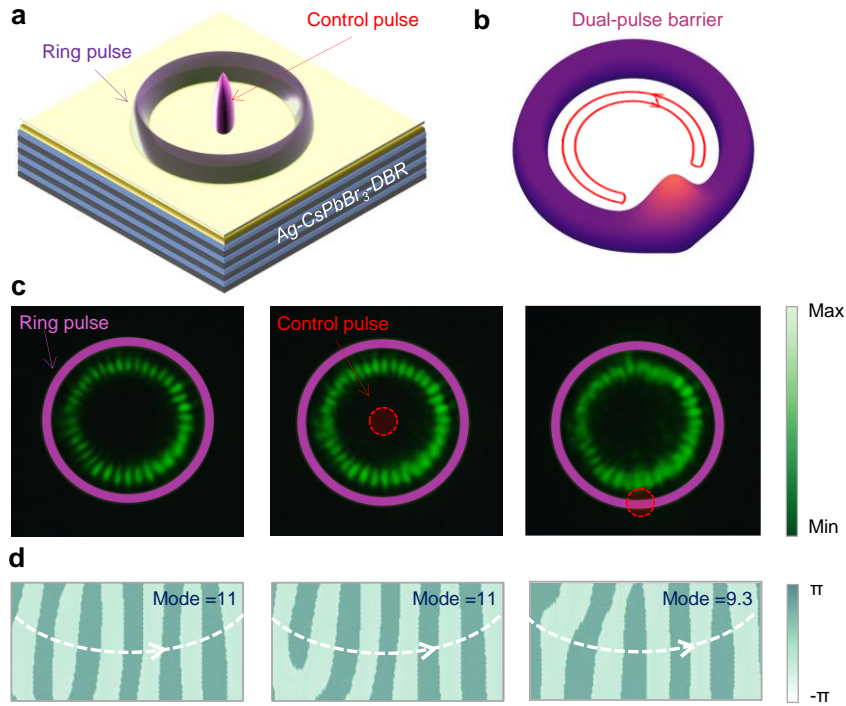

**Figure 2| All-optical symmetric control of high-order orbital modes.** (**a**) Schematic of spatial control of high-order OAM modes. On the basis of the annular pulse excitation above the threshold, an additional Gaussian-distributed control pulse is introduced, positioned at zero-point time and spatially adjustable. (**b**) The reformation of potential barrier as the control pulse is positioned on the ring pulse. The superposition of the two beams leads to the localized accumulation of excitons/exciton-polaritons, enabling the control of the EPs condensates. (**c**) The petal-shaped condensates in real-space images when control pulse at different positions. Purple line represents the ring pulses, and red line represents the control pulses. (**d**) Observing the changes of phase distribution in the fixed area of interest. The distribution of orbital modes corresponds to (**c**). The single annular excitation generates 11 modes in the area of interest (left part), and dual-pulse with cylindrical symmetry excited 11 modes with slightly edge expansion (middle part). When the cylindrical symmetry is broken, the mode number is switched to 9.3 (right part).

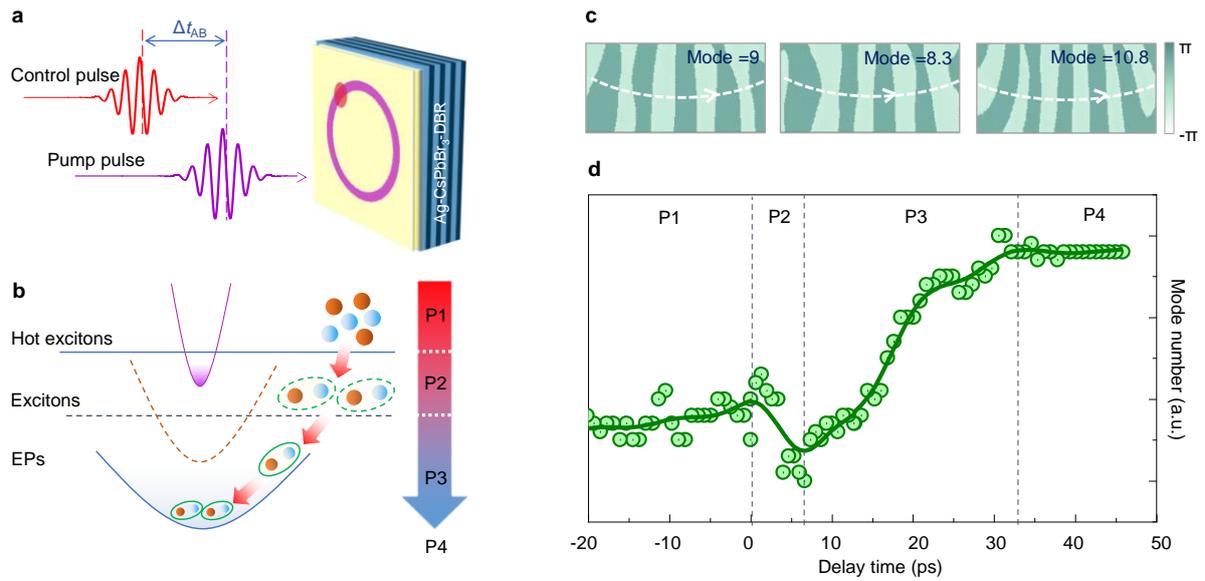

**Figure 3| All-optical temporal control of high-order orbital modes.** (**a**) Schematic of temporal control of high-order OAM modes. The control pulse is positioned on the ring pulse. By controlling the relative delay time ($\Delta t_{AB}$) of the two pulses, ultrafast control of the petal-shaped condensates is achieved. (**b**) The reservoir dynamics of condensation. Different delay times (P1, P2, P3, P4) regulate the mode emission by influencing distinct processes involved in the condensate dynamics. (**c**) The changes of phase distribution at different delay times. Corresponding to states at 0 ps, 5 ps, and 30 ps. (**d**) Curve of localized mode number variation with time delay. The gray dashed lines delineate regions corresponding to different dynamic processes.

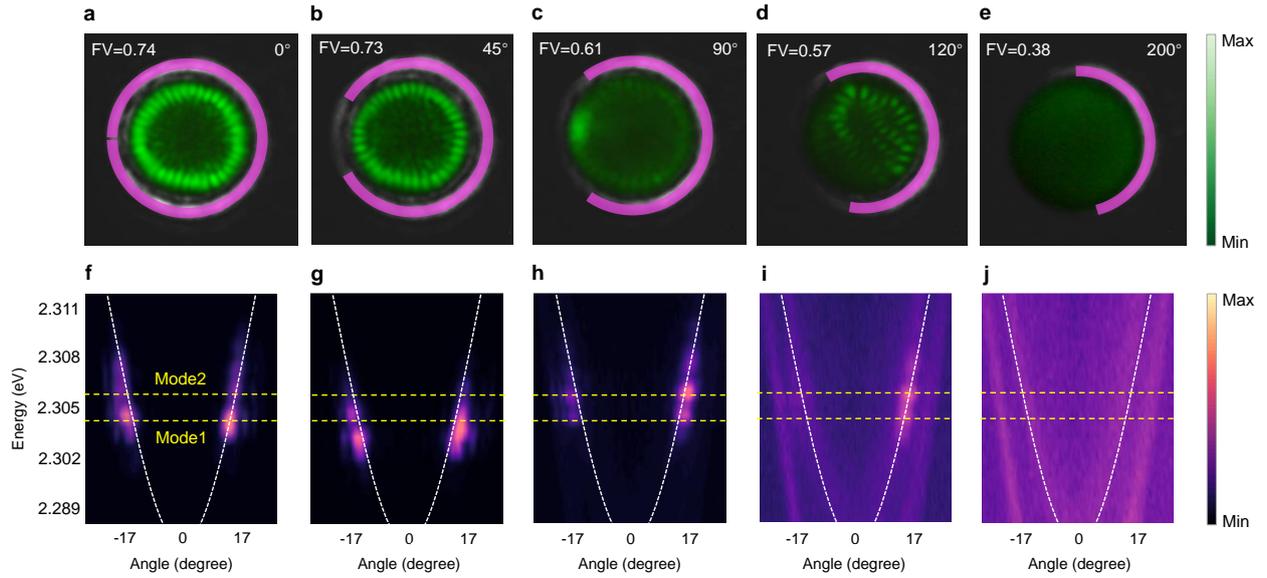

**Figure 4| the robustness verification of high-order orbital modes.** (**a-e**) The change of petal-shaped condensates in real-space images as the gap continues to increase. The gap opens from 0 ° to 200 °, and the fringe visibility of the stripes gradually decreases. Purple lines represent the ring pulses, with petal-shaped condensates in real-space images inside the ring and the background showing the intensity distribution of ring pulses. (**f-j**) Momentum-space imaging of the petal-shaped condensates when the gap is opened gradually. The white dashed line represents the LPB of EPs, whereas the yellow dashed line represents the two main orbital modes.

**TOC graphic**

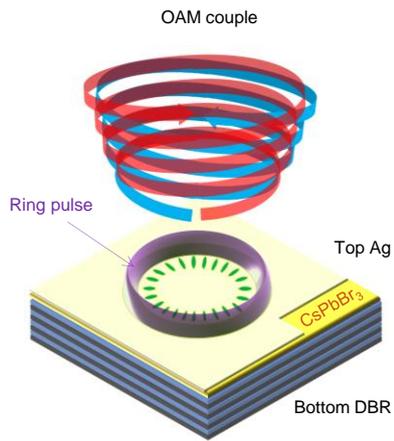
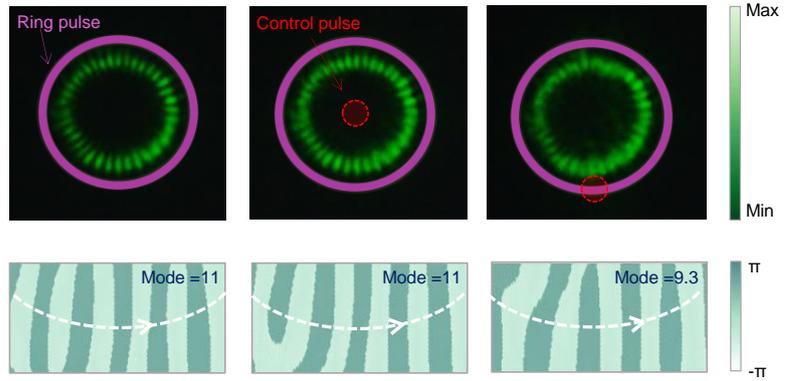